\documentclass[twocolumn,aps,prl,showpacs,groupedaddress,preprintnumbers,amsmath,amssymb,dvipdfmx]{revtex4}

\usepackage{graphicx}
\usepackage{pdfpages}
\usepackage{here}

\begin{document}

\title{
Fast parametric two-qubit gates with suppressed residual interaction \\using a parity-violated superconducting qubit
}

\author{Atsushi~Noguchi$^{1,2,3}$}
\email[]{u-atsushi@g.ecc.u-tokyo.ac.jp}
\author{Alto Osada$^{1,3}$}
\author{Shumpei~Masuda$^{4}$}
\author{Shingo~Kono$^{2,5}$}
\author{Kentaro~Heya$^{2}$}
\author{Samuel Piotr Wolski$^{2}$}
\author{Hiroki Takahashi$^{2}$}
\author{Takanori Sugiyama$^{2}$}
\author{Dany Lachance-Quirion$^{2}$}
\altaffiliation[Current address: ]{Nord Quantique, Sherbrooke, Qu\'{e}bec, J1K 0A5, Canada}
\author{Yasunobu~Nakamura$^{2,5}$}

\affiliation{%
$^{1}$Komaba Institute for Science (KIS), The University of Tokyo, Meguro-ku, Tokyo, 153-8902, Japan,\\
$^{2}$Research Center for Advanced Science and Technology (RCAST), The University of Tokyo, Meguro-ku, Tokyo, 153-8904, Japan,\\
$^{3}$PRESTO, Japan Science and Technology Agency, Kawaguchi-shi, Saitama 332-0012, Japan,\\
$^{4}$College of Liberal Arts and Sciences, Tokyo Medical and Dental University, Ichikawa, Chiba 272-0827, Japan,\\
$^{5}$Center for Emergent Matter Science (CEMS), RIKEN, Wako-shi, Saitama 351-0198, Japan
}

\date{\today}

\begin{abstract}
We demonstrate fast two-qubit gates using a parity-violated superconducting qubit consisting of a capacitively-shunted asymmetric Josephson-junction loop under a finite magnetic flux bias.
The second-order nonlinearity manifesting in the qubit enables the interaction with a neighboring single-junction transmon qubit via first-order inter-qubit sideband transitions with Rabi frequencies up to 30~MHz.
Simultaneously, the unwanted static longitudinal~(ZZ) interaction is eliminated with ac Stark shifts induced by a continuous microwave drive near-resonant to the sideband transitions.
The average fidelities of the two-qubit gates  are evaluated with randomized benchmarking as 0.967, 0.951, 0.956 for CZ, iSWAP and SWAP gates, respectively.
\end{abstract}

\maketitle

Quantum information processing with superconducting qubits has been intensively studied recently.
High-fidelity quantum manipulations and projective measurements have been achieved in multi-qubit systems~\cite{martinis2014,ibm2016, rigetti, martinis2019, johnson2020}, and basic quantum error-correction protocols have been demonstrated~\cite{schoelkopf2012, martinis2015, schoelkopf2016, correct2019, wallraf2019}.
For fault-tolerant quantum computing, however, the gate and readout fidelity should be further improved by a few orders of magnitude~\cite{fowler, terhal}.

To this end, a variety of two-qubit gates have been proposed and demonstrated.
These can be classified into two groups, based on their use of either a coupling between (near-)degenerate qubits~\cite{martinis2014, martinis2019} or a microwave-induced parametric coupling~\cite{nakamura2007, ibm2011, ibm2016, rigetti, johnson2020, screview}.
For the gate operation, the former usually requires fast frequency tuning of the qubits and/or a coupler through a flux bias, while the latter only uses microwave pulses for the dynamical control.
For the parametric gates, qubits are usually far off-resonant from each other in order to suppress residual couplings between them.
On the other hand, a large detuning slows down the parametric gate, causing a trade-off that hinders the improvement of the gate fidelity.
Recent works have addressed this issue by introducing various types of coupler circuits to eliminate the residual coupling without sacrificing the gate speed significantly~\cite{DiVincenzo2017, steele2018, houck2019,duan2020}. 
A simpler scheme combining two qubits with opposite signs of anharmonicity also allows a residual-coupling-free two-qubit cross-resonance gate~\cite{plourde2020}. 

In this letter, we propose and demonstrate fast parametric two-qubit gates using sideband transitions between an ordinary transmon qubit and a parity-violated superconducting qubit, which we call a cubic transmon.
The parity violation originates in a cubic component of the inductive potential of a Josephson-junction circuit under a finite magnetic flux bias.
This circuit, known as a superconducting nonlinear asymmetric inductive element~(SNAIL), was recently proposed~\cite{SNAIL} and utilized in parametric amplifiers~\cite{SNAIL2}, bosonic-mode qubits~\cite{SNAIL3}, and hybrid quantum systems~\cite{noguchi2018}. 
The parity symmetry breaking is essential for a physical system to acquire a second-order nonlinearity, which allows three-wave-mixing-type first-order sideband transitions and thus the parametric interactions with a neighboring qubit~\cite{nakamura2015,sideband, DiVincenzo2016}.
The large capacitive coupling strength between the qubits introduces strong parametric interactions, but also a large residual static interaction.
We eliminate the latter by using ac Stark shifts induced by a continuous near-resonant drive of the sideband transitions and solve the trade-off.
This approach of microwave-assisted elimination of the static interactions brings in more tuning knobs, i.e.~amplitudes and frequencies of multiple drives, which can be applied to the cases with multiple qubits and higher-order residual couplings. 

\begin{figure}[t]
\begin{center}
   \includegraphics[width=8.0cm,angle=0]{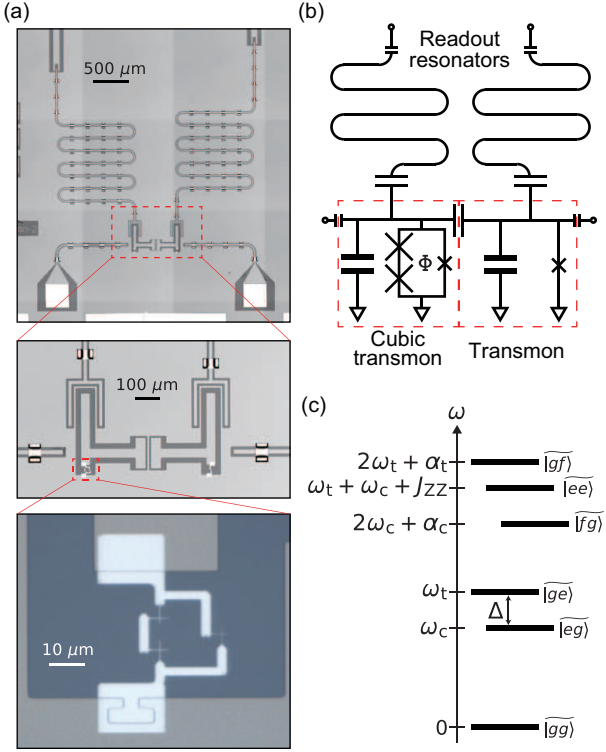}
\caption{
(a)~Photographs of the device. Most of the structures are made from Nb electrodes~(light gray) on a Si substrate~(dark gray), and the Josephson junctions~(at the three crosses in the bottom picture) are made of Al/AlO$_x$/Al junctions evaporated together with Al electrodes~(white). 
Air-bridges across the coplanar resonators and transmission lines suppress spurious modes on the chip.
(b)~Circuit diagram of the device.
(c)~Eigenstates $\widetilde{|ij\rangle}$~($ i,j \in \{ g,e,f\} $) of the two qubit system. The vertical axis indicates the eigenfrequency of the states.
}
\label{fig1}
\end{center}
\end{figure}
\begin{figure}[t]
\begin{center}
   \includegraphics[width=8.0cm,angle=0]{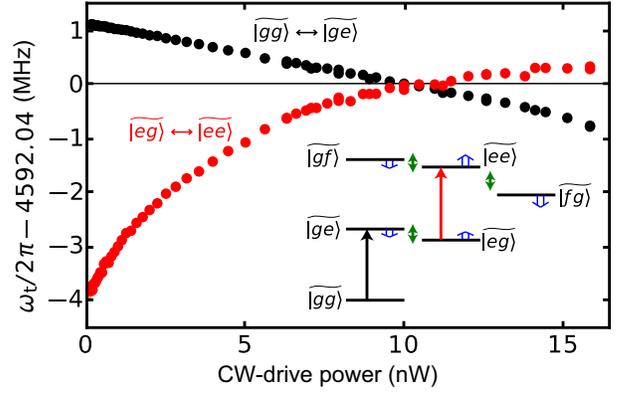}
\caption{
Elimination of the static ZZ interaction with a continuous-wave~(CW) drive.
The black and red dots are respectively the frequencies of the $\widetilde{| gg\rangle} \leftrightarrow \widetilde{| ge\rangle}$ and $\widetilde{| eg\rangle} \leftrightarrow \widetilde{| ee\rangle}$ transitions, determined by spectroscopy.
The inset shows the energy diagram.
The black and red arrows respectively indicate the corresponding transitions, and the green arrows represent the CW drive, which  simultaneously couples to all the sideband transitions. 
The blue arrows show the directions of the ac Stark shifts of the eigenstates.
The transition frequencies become identical at the CW-drive power of $\sim 10$~nW, and the static ZZ interaction is eliminated.
}
\label{fig2}
\end{center}
\end{figure}

Figures~1(a) and 1(b) present optical micrographs and a circuit diagram of the device, which contains two superconducting qubits and two resonators for the dispersive readout of each qubit.
The qubit on the right-hand side is a conventional transmon, which is composed of a capacitively-shunted single Josephson junction~\cite{koch2007}.
The other qubit is a cubic transmon, which is a capacitively-shunted SNAIL circuit.
The SNAIL is a Josephson-junction loop formed by a parallel circuit of a single small Josephson-junction and two large Josephson junctions. 
The SNAIL loop is threaded by a flux $\Phi$.
The Hamiltonian of the two-qubit system reads
\begin{eqnarray}
\hat{H}/\hbar =\!\!\!\!\!&&\omega _\mathrm{c0}\hat{a}^\dagger\hat{a}+\beta_\mathrm{c0} (\hat{a}^\dagger\hat{a}^\dagger\hat{a}+\hat{a}^\dagger\hat{a}\hat{a})+\frac{\alpha_\mathrm{c0}}{2}\hat{a}^\dagger\hat{a}^\dagger\hat{a}\hat{a}\nonumber\\
&&+\omega _\mathrm{t0}\hat{b}^\dagger\hat{b}+\frac{\alpha_\mathrm{t0}}{2}\hat{b}^\dagger\hat{b}^\dagger\hat{b}\hat{b}
+g_0(\hat{a}^\dagger\hat{b}+\hat{a}\hat{b}^\dagger ),
\end{eqnarray}
where $\hbar = h/2\pi $ is the reduced Planck constant, $\omega _\mathrm{c0}$ and $\omega _\mathrm{t0}$ are the bare eigenmode frequencies, and $\hat{a}$ and $\hat{b}$ are the annihilation operators for the cubic transmon and conventional transmon, respectively.
The coefficient $\beta_\mathrm{c0}$ is the second-order nonlinearity of the cubic transmon, $\alpha _\mathrm{c0}$ and $\alpha _\mathrm{t0}$ are the third-order nonlinearities of each qubit, and $g_0$ is the capacitive coupling strength between the two qubits.

In the dispersive coupling regime $|\Delta _0 |\equiv |\omega _\mathrm{c0}-\omega _\mathrm{t0}|\gg g_0$,
the effective Hamiltonian can be written as
\begin{eqnarray}
\hat{H}_\mathrm{eff}/\hbar =\!\!\!\!\!&&\left[ \omega _\mathrm{c}+g(\hat{b}^\dagger +\hat{b})\right] \hat{a}^\dagger\hat{a}+\frac{\alpha_\mathrm{c}}{2}\hat{a}^\dagger\hat{a}^\dagger\hat{a}\hat{a}\nonumber\\
&&+\omega _\mathrm{t}\hat{b}^\dagger\hat{b}+\frac{\alpha_\mathrm{t}}{2}\hat{b}^\dagger\hat{b}^\dagger\hat{b}\hat{b}+J_\mathrm{ZZ}\hat{a}^\dagger\hat{a}\hat{b}^\dagger\hat{b},
\end{eqnarray}
where $g$($\propto \beta_\mathrm{c0}$) is the effective coupling strength, and $\omega_\mathrm{c}$, $\omega_\mathrm{t}$, $\alpha_\mathrm{c}$, and $\alpha_\mathrm{t}$ are the eigenmode frequencies and self-Kerr nonlinearities of the qubits after the perturbative treatment of the coupling term in Eq.~(1), respectively. 
The term with a coefficient $g$ arises from the second-order nonlinearity in the parity-violated cubic transmon and gives the interaction in the same form as the radiation pressure in optomechanics~\cite{optomechanics, noguchi2018} and the state-dependent force in trapped ions~\cite{blatt2008,ion}.
There is also a static longitudinal~(ZZ) interaction between the qubits, whose amplitude is $J_\mathrm{ZZ}$.
The detailed derivations and expressions of the parameters in Eqs.~(1) and~(2) are presented in the Supplementary Material~\cite{supple}.

Figure~1(c) illustrates the eigenstates $\widetilde{|ij\rangle}$~$( i,j \in \{ g,e,f \})$ and their frequencies.
The coupling between qubits hybridizes the bare qubit states, $|i\rangle_\mathrm{c} |j\rangle_\mathrm{t}$, and forms the eigenstates $\widetilde{|ij\rangle}$~\cite{supple}.
When a drive field at frequency $\omega_\mathrm{d}=\Delta \equiv \omega_\mathrm{t} - \omega_\mathrm{c}$ is applied to the cubic transmon, the two qubits resonate with each other in the rotating frame.
Under the rotating-wave approximation, the parametric coupling follows
\begin{eqnarray}
\hat{H}_\mathrm{p}/\hbar \!\!&=&
\!\! \eta\Omega (e^{i\omega _\mathrm{d}t+i\theta}\hat{a}^\dagger \hat{b}+e^{-i\omega _\mathrm{d}t-i\theta}\hat{a} \hat{b}^\dagger ),\\
\eta \!\!&\equiv &\!\! \frac{-2g_0\beta_\mathrm{c0}(2\omega _\mathrm{c0}^2-\alpha_\mathrm{c0}\Delta _0+ 2\alpha_\mathrm{c0}\omega_\mathrm{c0})}{\Delta _0 (\Delta _0 -\omega _\mathrm{c0})(\alpha_\mathrm{c0}+\omega _\mathrm{c0})(\alpha_\mathrm{c0}+\omega _\mathrm{c0}+\Delta _0)},~
\end{eqnarray}
where $\Omega$ and $\theta$ are the amplitude and phase of the drive field, respectively.
Note that $\eta$ is proportional to the second-order nonlinearity $\beta _\mathrm{c0}$ for the three-wave-mixing process occurring under the Hamiltonian $\hat{H}_\mathrm{p}$ in Eq.~(3). 

\begin{figure*}[t]
\begin{center}
   \includegraphics[width=15.4cm,angle=0]{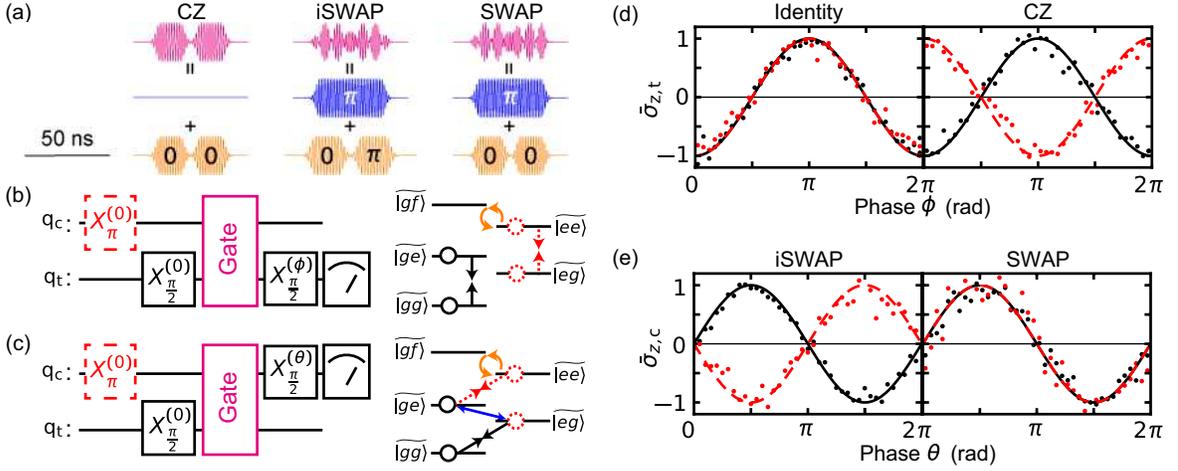}
\caption{
Implementation of the two-qubit gates.
(a) Composite pulses for CZ, iSWAP and SWAP gates, applied to the cubic transmon.
The frequencies of the blue and orange pulses are resonant to the $\widetilde{|ge\rangle} \leftrightarrow \widetilde{|eg\rangle}$ and $\widetilde{|ee\rangle} \leftrightarrow \widetilde{|gf\rangle}$ transitions, respectively.
The relative phases of the second segment of the orange pulses are fine-tuned to optimize the amount of the conditional phase.
(b)~[(c)]~Pulse sequence for characterizing the two-qubit gates, identity and CZ~[iSWAP and SWAP], and corresponding transitions.
$X_\tau^{(\phi)}$~($X_\tau^{(\theta)}$) represents a single-qubit rotation with the rotation angle $\tau$ and phase $\phi$~($\theta$).
The right panels in (b) and (c) are energy-level diagrams illustrating the Ramsey interferometries involving a two-qubit gate. 
The black solid and red dashed arrows indicate pairs of interfering eigenstates. 
The blue and orange arrows indicate the sideband transitions. 
(d)~[(e)]~Ramsey-type interference, conditioned on the state of the cubic transmon, with the two-qubit gates such as identity and CZ~(iSWAP and SWAP) gates.
The excitation probability of the transmon, $\bar{\sigma}_{z,\mathrm{t}}$~(the cubic transmon, $\bar{\sigma}_{z,\mathrm{c}}$) is determined by the average readout in the time-ensemble measurement. 
The black (red) dots show the experimental results without~(with) the initial $X_\pi$-pulse [red dashed rectangles in (b) and (c)] for the cubic transmon. 
The black solid and red dashed curves represent the functions for the ideal gates.
}
\label{fig3}
\end{center}
\end{figure*}

Under the resonant condition $\omega _\mathrm{d} = \Delta$, the drive exchanges the excitation of the two qubits, and thus the iSWAP and SWAP gates can be implemented. 
Another type of two-qubit gate, controlled-phase (CZ) gate, is similarly achieved with a parametric drive.
When the drive frequency is equal to $\Delta + \alpha _\mathrm{t}$, the transition $\widetilde{|ee\rangle} \leftrightarrow \widetilde{|gf\rangle}$ takes place. 
A $2\pi$-rotation of the transition induces a geometric phase factor of $-1$ only to the $\widetilde{|ee\rangle}$ state in the computational subspace.

In parallel with the dynamically-induced coupling, there remains the spurious static ZZ interaction, the last term in Eq.~(2), between the capacitively coupled qubits with higher energy levels~\cite{ibm2019}.
Remarkably, the residual interaction can be eliminated also with a parametric drive.
We irradiate the cubic transmon with a continuous-wave (CW) microwave field, whose frequency is slightly detuned from the transition of $\widetilde{|ee\rangle} \leftrightarrow \widetilde{|fg\rangle}$.
The ac Stark effect by the CW drive shifts the eigenfrequencies in the two-qubit subspace.
These shifts give rise to a tunable ZZ interaction and allow compensation for the unwanted interaction.

In the experiment, we use the device shown in Fig.~1.
The parameters at the operating flux bias, $\Phi =0.34\Phi _0$, where $\Phi _0\equiv h/2e$, are the followings:
The eigenfrequencies of the cubic transmon and the transmon are $\omega _\mathrm{c}/2\pi = 3.633\mathrm{~GHz}$ and $\omega _\mathrm{t}/2\pi = 4.479\mathrm{~GHz}$, respectively.
The third-order nonlinearities of the qubits are $\alpha _\mathrm{c}/2\pi = -132~\mathrm{MHz}$ and $\alpha _\mathrm{t}/2\pi = -168~\mathrm{MHz}$, and the bare coupling strength between the qubits is $g_0/2\pi = 75 \mathrm{~MHz}$, which are determined by spectroscopic measurements.
The details of the sample characterization are described in the Supplementary Material~\cite{supple}.
Using these values, we estimate the second-order nonlinearity $\beta_\mathrm{c} /2\pi = -195 \mathrm{~MHz}$, the effective coupling strength $g/2\pi = -14\mathrm{~MHz}$, and the coupling coefficient of the parametric drive, $\eta = 0.022$. 
The energy-relaxation and Ramsey-dephasing times of the qubits are $T_1 = 3.9 \mathrm{~\mu s}$ and $T_2^* = 0.6 \mathrm{~\mu s}$ for the cubic transmon, and $T_1 = 4.0 \mathrm{~\mu s}$ and $T_2^* = 2.3 \mathrm{~\mu s}$ for the transmon, respectively.
The dephasing time of the cubic transmon is improved to $T_2^\mathrm{E} = 1.5\mathrm{~\mu s}$ with an echo pulse, while no change is seen for the transmon.

We first eliminate the residual ZZ interaction by the CW drive (Fig.~2).
The drive frequency is 930 MHz, and the detuning from the $\widetilde{|ge\rangle} \leftrightarrow \widetilde{|eg\rangle}$ transition is 84~MHz.
The inset of Fig.~2 shows the shifts of eigenstates induced by the CW drive. 
The CW drive is red detuned from the $\widetilde{|ee\rangle} \leftrightarrow \widetilde{|fg\rangle}$ transition and blue detuned from the $\widetilde{|ge\rangle} \leftrightarrow \widetilde{|eg\rangle}$ and $\widetilde{|ee\rangle} \leftrightarrow \widetilde{|gf\rangle}$ transitions.
Thus, the sign of the frequency shift is different from each other. 
The amplitude of the ZZ interaction corresponds to the frequency difference between the $\widetilde{| gg\rangle} \leftrightarrow \widetilde{| ge\rangle}$ and $\widetilde{| eg\rangle} \leftrightarrow \widetilde{| ee\rangle}$ transitions, which amounts to $5$~MHz in the absence of the CW drive.
The frequency difference vanishes at a certain power of the drive.
It is also found that the CW drive does not degrade the coherence of the qubits~(data not shown).

In the presence of the CW drive, we implement the two-qubit Clifford gate set, i.e.~CZ, iSWAP and SWAP gates, using parametric couplings induced by additional microwave pulses.
These gates are within the family of the Fermionic Simulation gate set, characterized by two parameters, the swap angle $\theta _\mathrm{sw}$ and the conditional phase $\theta _\mathrm{cp}$~\cite{fermi2018, martinis2020}.
The CZ, iSWAP, and SWAP gates have the parameters $(\theta _\mathrm{sw} ,\theta _\mathrm{cp} )=(0, \pi)$, $(\pi , 0)$ and $(\pi , \pi)$, respectively.
In our setup, $\theta_\mathrm{sw}$ and $\theta_\mathrm{cp}$ are independently and simultaneously controlled via parametric couplings.

Figure~3(a) illustrates the waveforms of the synthesized two-tone pulses for these gates. 
The total gate time is 50~ns for each.
The swap pulse (blue) is resonant to the $\widetilde{|ge\rangle} \leftrightarrow \widetilde{|eg\rangle}$ transition and is used to control $\theta_\mathrm{sw}$. 
The control-phase pulse (orange) is resonant to the $\widetilde{|ee\rangle} \leftrightarrow \widetilde{|gf\rangle}$ transition and controls $\theta_\mathrm{cp}$ through the relative phase between two serial segments, applying the conditional phase $\theta_\mathrm{cp}$ as a geometrical phase only to the $\widetilde{|ee\rangle}$ state. 
Because the swap pulses also generate a small conditional phase due to the Stark shift, we simultaneously apply a control-phase pulse for iSWAP and SWAP gate to eliminate the unwanted phase.

Figure 3(b) [3(c)] shows the pulse sequence of the Ramsey interferometry for characterizing the identity and CZ gates [iSWAP and SWAP gates].
For the iSWAP and SWAP gates, which exchange an excitation between the qubits, we apply the second $\pi$/2-pulse to the cubic transmon instead of the transmon to form an interferometric sequence~[Fig.~3(c)], in contrast to the standard Ramsey experiments.
Figures 3(d) and 3(e) show the experimental data of Ramsey oscillations, conditioned on the state of the cubic transmon, revealing the amount of the conditional phase $\theta _\mathrm{cp}$ of each two-qubit gate.
The phase difference between the Ramsey oscillations, with and without an initial $\pi$-rotation of the cubic transmon, corresponds to the conditional phase shift.
The experimental data have good agreements with the ideal behaviors in Figs.~3(d) and~3(e). 

\begin{figure}[t]
\begin{center}
   \includegraphics[width=7.0cm,angle=0]{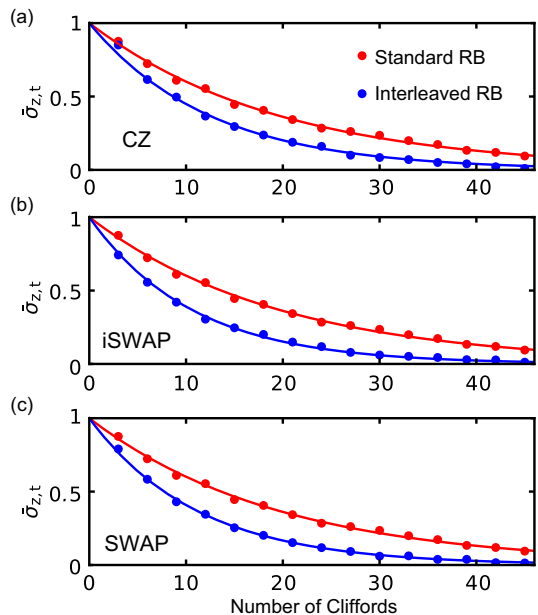}
\caption{
Randomized benchmarking~(RB) for two-qubit gates.
The vertical axes show the normalized average quadrature amplitude $\bar{\sigma}_{z,\mathrm{t}}$ of the transmon readout signal~\cite{RB2}.
The horizontal axes show the number of Clifford gates in the randomized sequence. 
The red dots show the result of standard RB with two-qubit gates, and the blue dots express those of interleaved RB for (a) CZ, (b) iSWAP and (c) SWAP gates, respectively.
}
\label{fig4}
\end{center}
\end{figure}

Finally, we characterize the average two-qubit gate fidelities with the randomized-benchmarking (RB) protocols~\cite{RBth1, RBth2, RB1}.
The gate time is uniformly set to 50 ns for the CZ, iSWAP, SWAP gates and all the single-qubit gates.
Figure 4 shows the experimental results of the two-qubit RB.
From the standard RB, the average gate fidelity of the two-qubit Clifford gates is determined to be $0.950 \pm 0.001$. 
Using this value and those from the interleaved RB, we estimate the average gate fidelity of each two-qubit gate: $0.971 \pm 0.002$, $0.958 \pm 0.001$ and $0.962 \pm 0.001$ for CZ, iSWAP and SWAP gates, respectively.
The achieved fidelities of the two-qubit gates are comparable to those of the single-qubit gates~\cite{supple} and mostly limited by the energy relaxation time of the qubits.
The coherence limits are approximately 0.97 according to the gate pulse widths, which are close to the
observed fidelities.
As the cubic transmon has basically the same layout as conventional transmons, we expect improvement of the relaxation time through optimizations of the design and fabrication.

In conclusion, we have demonstrated a parity-violated qubit called a cubic transmon, and realized microwave-controlled fast two-qubit gates between a cubic transmon and a conventional transmon.
As the gates originate from the second-order nonlinearity of the circuit, the coupling strength scales inversely proportional to the detuning between the qubits, not to the square of it, and is thus sufficiently large for a wide detuning range of the qubits. 
This is advantageous for a multi-qubit system, which often suffers from a frequency-crowding problem.
The residual static ZZ interaction is eliminated by applying a continuous microwave field, which will allow us to increase the bare coupling strength further and make the two-qubit gates as fast as 20~ns with optimal device parameters.
This scheme for the suppression of the residual coupling can also be extended to multi-qubit systems as well as to higher-order interactions by using multiple drives.

The authors acknowledge Y. Sunada, K. Nittoh and K. Kusuyama for the help in sample fabrication and W. Oliver for providing a TWPA.
This work was partly supported by JSPS KAKENHI (Grant Number 26220601, 18K03486), JST PRESTO (Grant Number JPMJPR1429), JST ERATO (Grant Number JPMJER1601), and Q-LEAP (Grant Number JPMXS0118068682).

　 
\widetext
\clearpage

\begin{center}
\textbf{\large Supplementary Material for\\
``Fast parametric two-qubit gates with suppressed residual interaction \\using a parity-violated superconducting qubit''}
\end{center}
\setcounter{equation}{0}
\setcounter{figure}{0}
\setcounter{table}{0}
\setcounter{page}{1}
\makeatletter
\renewcommand{\bibnumfmt}[1]{[S#1]}
\renewcommand{\citenumfont}[1]{S#1}

\renewcommand{\thetable}{S\arabic{table}}
\renewcommand{\thefigure}{S\arabic{figure}}
\renewcommand{\theequation}{S\arabic{equation}}

\begin{figure}[h]
\begin{center}
   \includegraphics[width=14.0cm,angle=0]{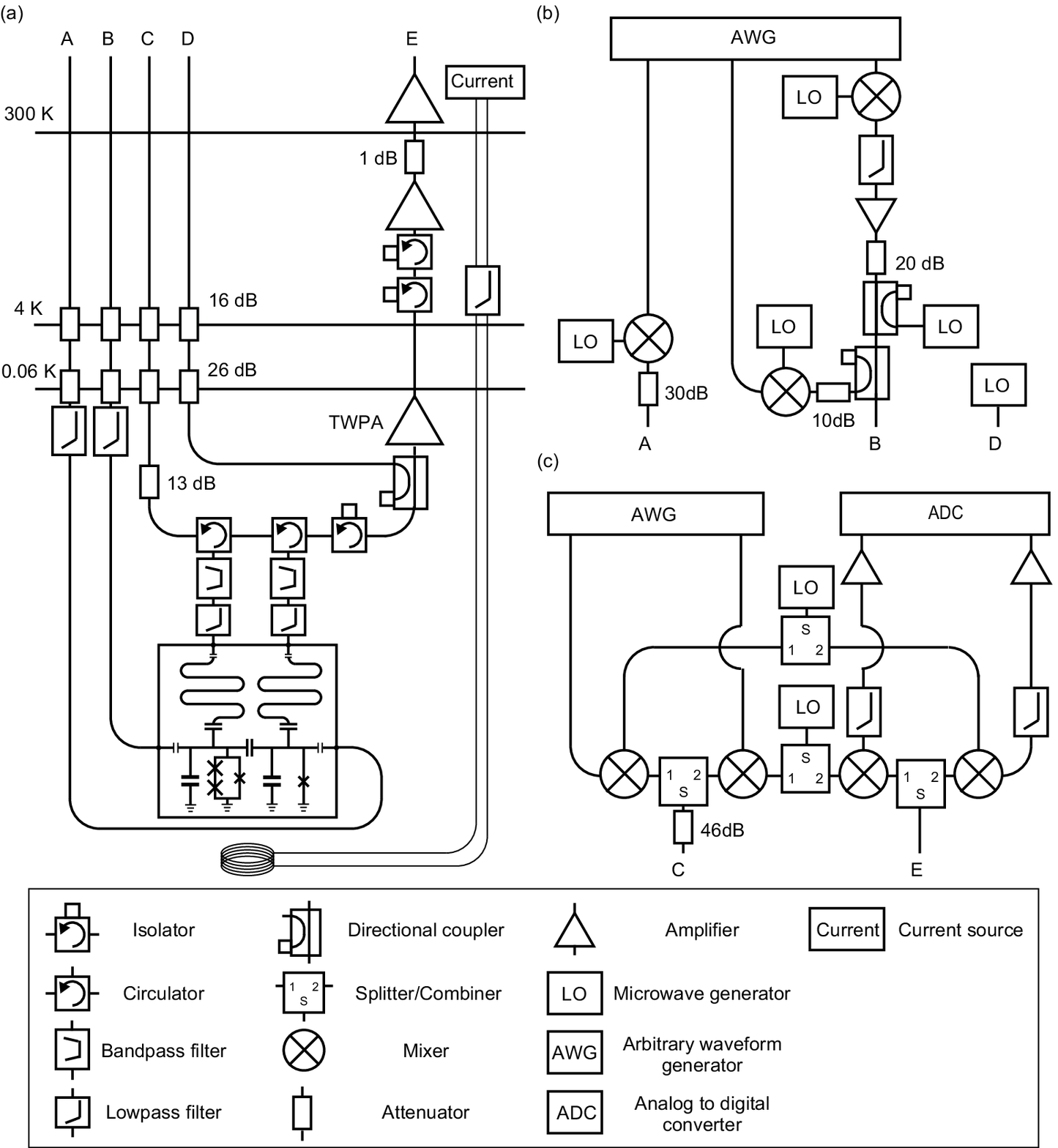}
\caption{
Wiring scheme of the experimental setup. 
(a)~Connections from the sample chip to the ports at room temperature.
(b)~Pulse generating system for the qubit control.
(c)~Readout system.
All the local oscillators (LOs) in~(b) and~(c) are phase-locked with a 10-MHz reference clock.
}
\label{figSwire}
\end{center}
\end{figure}

\section{Measurement setup}
Figure S1 illustrates the wiring scheme for the gate experiments.
The sample chip is connected to one readout port~(C) and two drive ports (A, B).
We apply microwave pulses generated by modulating the local oscillator signals.
The qubits are simultaneously read out with the dispersive technique.
The resonance frequencies and total decay rates of the readout resonators are 6.767~GHz and 0.8~MHz for the cubic transmon and  6.509~GHz and 1.0~MHz for the transmon, respectively.
The reflection signals of the readout resonators are amplified by a Josephson traveling wave parametric amplifier (TWPA) and two  low-noise amplifiers and demodulated for the readout.
We apply a magnetic flux into the SNAIL loop through an external superconducting coil.

\section{SINGLE-PHASE APPROXIMATION OF CUBIC TRANSMON}
Figure~S2 shows the full-circuit model of a cubic transmon.
Each of the two isolated superconducting islands has two degrees of freedom of the phase and charge.
The full Hamiltonian is written as
\begin{eqnarray}
H &=& K -E_\mathrm{J1} \cos \phi _1 - E_\mathrm{J2} \cos \phi _2 -E_\mathrm{J3} \cos (\phi + \phi _1-\phi _2 ),\\
K&=& 4 E_\mathrm{C} \, \left(n_1, n_2\right) \textbf{C}
\left(
  \begin{array}{c}
     n_1 \\
     n_2
  \end{array}
 \right)
,\\
\textbf{C}&=&
 \left(
  \begin{array}{cc}
     1+k_1+k_3 & -k_3 \\
     -k_3 & k_2+k_3
  \end{array}
 \right)^{\! -1},
\end{eqnarray}
where $n_1$ and $n_2$ are the numbers of excess Cooper pairs on each island, $\phi _1$ and $\phi _2$ are the superconducting phases across each junction connected to the ground, $\phi =2\pi\Phi /\Phi _0$ is the reduced magnetic flux, $\Phi$ is the flux threading the loop, $\Phi _0=h/2e$ is the flux quantum, and $k_i$~($i=1,2,3$) are the scaling factors depending on the junction size. 
$E_\mathrm{C}$ is the single-electron charging energy of the shunt capacitance.
We assume that each Josephson energy $E_{\mathrm{J}i}$ scales as $E_{\mathrm{J}i} = k_i E_\mathrm{J0}$, where $E_\mathrm{J0}$ is an independent parameter to be determined.

\begin{figure}[h]
\begin{center}
   \includegraphics[width=7.0cm,angle=0]{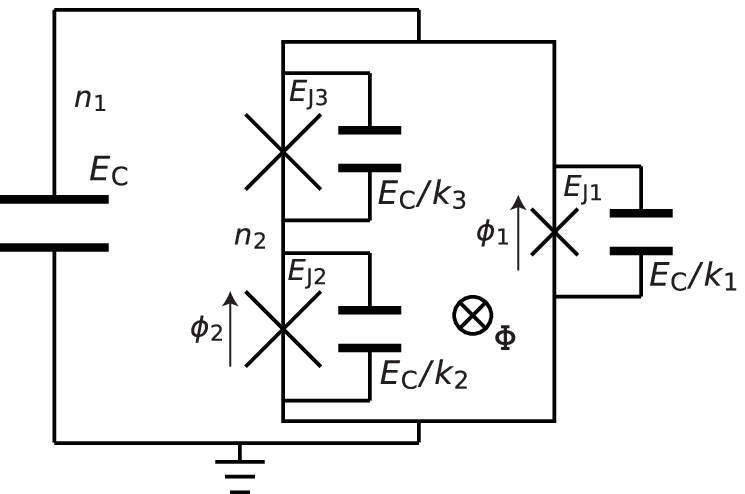}
\caption{
Full-circuit model of a cubic transmon.
}
\label{figScubic}
\end{center}
\end{figure}

\begin{figure}[t]
\begin{center}
   \includegraphics[width=15.0cm,angle=0]{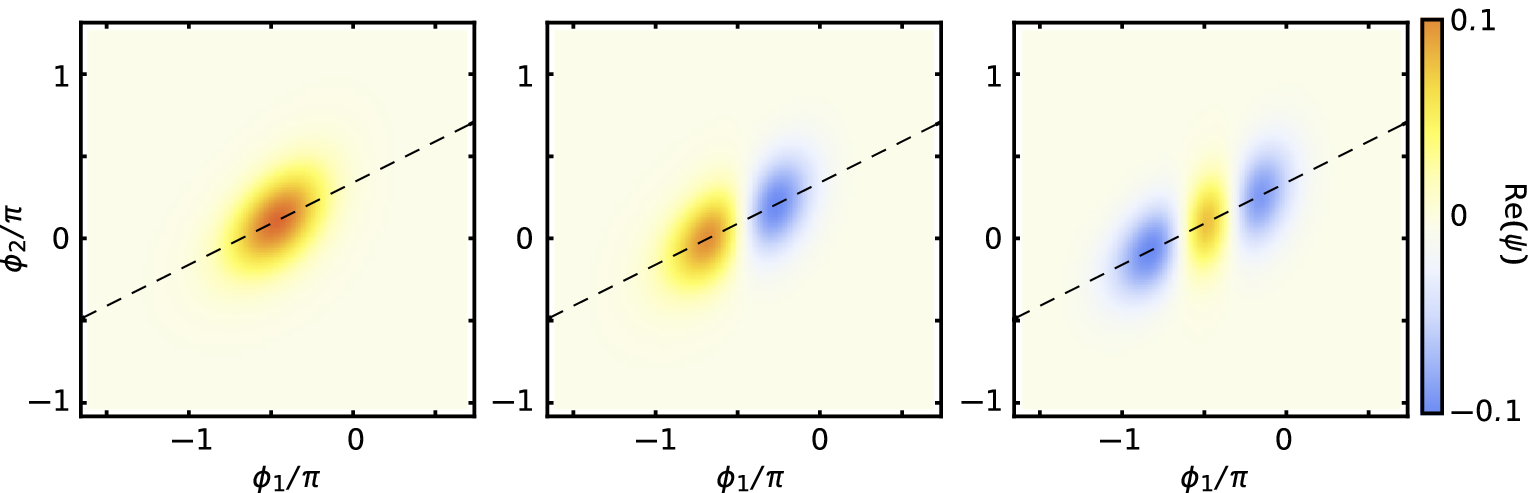}
\caption{
Wave functions of the cubic-transmon eigenstates, obtained by diagonalizing the Hamiltonian~[Eq.~(S1)] with the parameters in the second line of Table~S1, where a condition $k_2=k_3$ is assumed.
Real parts of the wave functions for the ground, first-excited and second-excited states are plotted from left to right, respectively.
The dashed lines represent the constraint $2\phi_2=\phi+\phi_1$, which is used in the single-phase approximation.
}
\label{figSwave}
\end{center}
\end{figure}

By diagonalizing the Hamiltonian, we obtain wave functions of the eigenstates of the cubic transmon in the phase representation~(Fig.~\ref{figSwave}).
Under the condition of $k_2=k_3$, the fringes of the excited states lie on the dashed lines indicating the relation $2 \phi _2 = \phi + \phi _1$.
The confinement of the wave functions along the dashed line suggests an approximation, $2 \phi _2 \approx \phi + \phi _1$, which we call the single-phase approximation.
Using this relation, we can write the inductive energy of the SNAIL, 
\begin{eqnarray}
U(\varphi )&=&-k_1E_\mathrm{J0}\cos \varphi -2k_2E_\mathrm{J0}\cos \left( \frac{\phi-\varphi}{2}\right)\nonumber\\
&=&D_2\delta ^2+D_3\delta ^3+D_4\delta ^4+O(\delta ^5).\label{eqpot}
\end{eqnarray}
This gives an effective model with a single phase degree of freedom, $\varphi \, (\equiv \phi_1)$. 
The Josephson energies in the main text are defined as $E_\mathrm{J}^\prime = k_1E_\mathrm{J0}$ and $E_\mathrm{J}=k_2E_\mathrm{J0}$.
The second formula in Eq.~(\ref{eqpot}) is the Taylor expansion around the phase $\varphi_0$ at a minimum of the inductive energy, where $\delta \equiv \varphi - \varphi_0$ is the relative phase variable for the expansion and $D_i$~$(i=2,3,4)$ are the expansion coefficients.
The parity symmetry $\delta \leftrightarrow -\delta$ is broken as seen in the existence of the $\delta ^3 $ term in the presence of a finite magnetic flux penetrating through the SNAIL loop.

Under the approximation, the Hamiltonian of the cubic transmon up to the third-order nonlinearity is, as in the main text,
\begin{eqnarray}
\hat{H}/\hbar =\!\!\!\!\!&&\omega _\mathrm{c0}\hat{a}^\dagger\hat{a}+\beta_\mathrm{c0} (\hat{a}^\dagger\hat{a}^\dagger\hat{a}+\hat{a}^\dagger\hat{a}\hat{a})+\frac{\alpha_\mathrm{c0}}{2}\hat{a}^\dagger\hat{a}^\dagger\hat{a}\hat{a},
\end{eqnarray}
where
\begin{eqnarray}
\hbar\omega _\mathrm{c0} &=& \sqrt{16 D_2 E_\mathrm{C}^\prime}+\frac{12D_4E_\mathrm{C}^\prime}{D_2}\\
\hbar\beta _\mathrm{c0} &=& 3\left( \frac{E_\mathrm{C}^\prime}{D_2}\right) ^{3/4} D_3 \\
\hbar\alpha _\mathrm{c0} &=& \frac{6D_4E_\mathrm{C}^\prime}{D_2}.
\end{eqnarray}
The effective charging energy $E_\mathrm{C}^\prime$ is expressed as
\begin{equation}
E_\mathrm{C}^\prime \equiv E_\mathrm{C} \frac{k_2+k_3}{k_2+k_3 + k_1 (k_2+k_3)+k_2k_3}.
\end{equation}

\begin{figure}[b]
\begin{center}
   \includegraphics[width=16.0cm,angle=0]{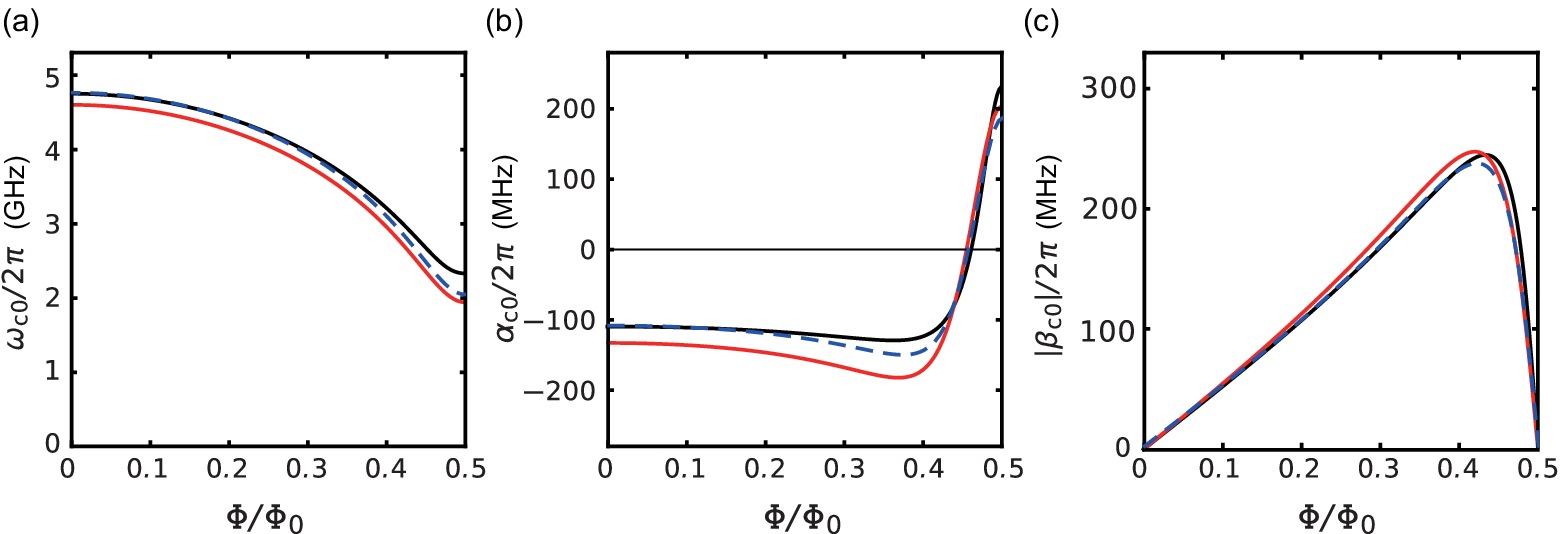}
\caption{
Accuracy of the single-phase approximation.
(a)~Eigenfrequency~$\omega_\mathrm{c0}$, (b)~third-order nonlinearity~$\alpha_\mathrm{c0}$ and (c)~second-order nonlinearity~$|\beta_\mathrm{c0}|$ of a cubic transmon as a function of $\Phi$.
Black curves show the calculations based on the single-phase approximation with the parameters determined by the fittings in Figs.~S5(a) and~S5(b).
Red curves are the simulation from the full-circuit model with the same parameters.
Blue dashed curves are from the full-circuit model with adjusted parameters.
The parameters are listed in Table~S1.
}
\label{figScomp}
\end{center}
\end{figure}

\begin{table}[b]
\caption{Parameters used for the calculations in Figs.~\ref{figSwave} and \ref{figScomp}. The values for the single-phase approximation are determined from the fittings in Figs.~\ref{figSfit}(a) and~\ref{figSfit}(b). Those for the full-circuit model are adjusted to obtain the blue dashed curves in Fig.~\ref{figScomp}, which closely reproduce the calculations based on the single-phase approximation.
}
\begin{ruledtabular}
\begin{tabular}{ccccc}
&$E_\mathrm{C}/h\mathrm{~(GHz)}$&$E_\mathrm{J0}/h\mathrm{~(GHz)}$&$k_1$&$k_2 (=k_3)$\\\hline
Single-phase approximation&0.21&84&0.070&0.20\\
Full-circuit model&0.18&103&0.070&0.20
\end{tabular}
\end{ruledtabular}
\end{table}

We quantitatively compare the single-phase approximation with the full-circuit model.
We calculate the eigenmode frequency of the first excited state, $\omega_\mathrm{c0}$, third-order nonlinearity $\alpha_\mathrm{c0}$ and second-order nonlinearity $\beta_\mathrm{c0}$ of the cubic transmon based on each model~(Fig.~\ref{figScomp}).
For the calculation with the single-phase approximation, we use the parameters obtained by the fittings in Figs.~\ref{figSfit}(a) and (b) below.
Next, we use the same parameters in the full-circuit model and compare the results~(red lines).
For the full-circuit model, the second-order nonlinearity $\beta _\mathrm{c0}$ is evaluated from the transition moment between the ground and second-excited states.
The transition moment is defined as
\begin{equation}
A_{ij}= |\langle i | n_i | j \rangle |,
\end{equation}
where $i, j \in \{ g, e, f \}$.
We also obtain from the perturbative approach
\begin{eqnarray}
A_{ge} &\propto& \Omega _0 \\
A_{gf} &\propto& \frac{2\beta_\mathrm{c0}\Omega _0}{\omega _\mathrm{c0}+\alpha_\mathrm{c0}},
\end{eqnarray}
where $\Omega _0$ is the external drive amplitude for these transition.
Using these, we calculate the absolute value of $\beta_\mathrm{c0}$ as
\begin{equation}
|\beta _\mathrm{c0}|= \frac{\omega _\mathrm{c0} +\alpha _\mathrm{c0}}{2} \frac{A_{gf}}{A_{ge}}.
\end{equation}

The calculations based on  these models qualitatively agree with each other [Fig.~\ref{figScomp}(a)--(c)] and demonstrate the validity and accuracy of the single-phase approximation.
There is a small quantitative deviation between the two models, which is not surprising as the wave functions of the eigenstates (Fig.~\ref{figSwave}) are not completely localized along the dashed line.
This means that $2\phi_2=\phi+\phi_1$ is not strictly satisfied because of the quantum fluctuation of $\phi _2$, and we cannot construct an exact single-phase model.
Blue dashed curves in Fig.~\ref{figScomp} show calculations based on the full-circuit model with adjusted parameters to reproduce the results of the single-phase approximation.
For the region with small reduced magnetic flux, these calculations have a good agreement with each other. 

\section{COUPLED QUBITS}
\begin{figure}[b]
\begin{center}
   \includegraphics[width=15.0cm,angle=0]{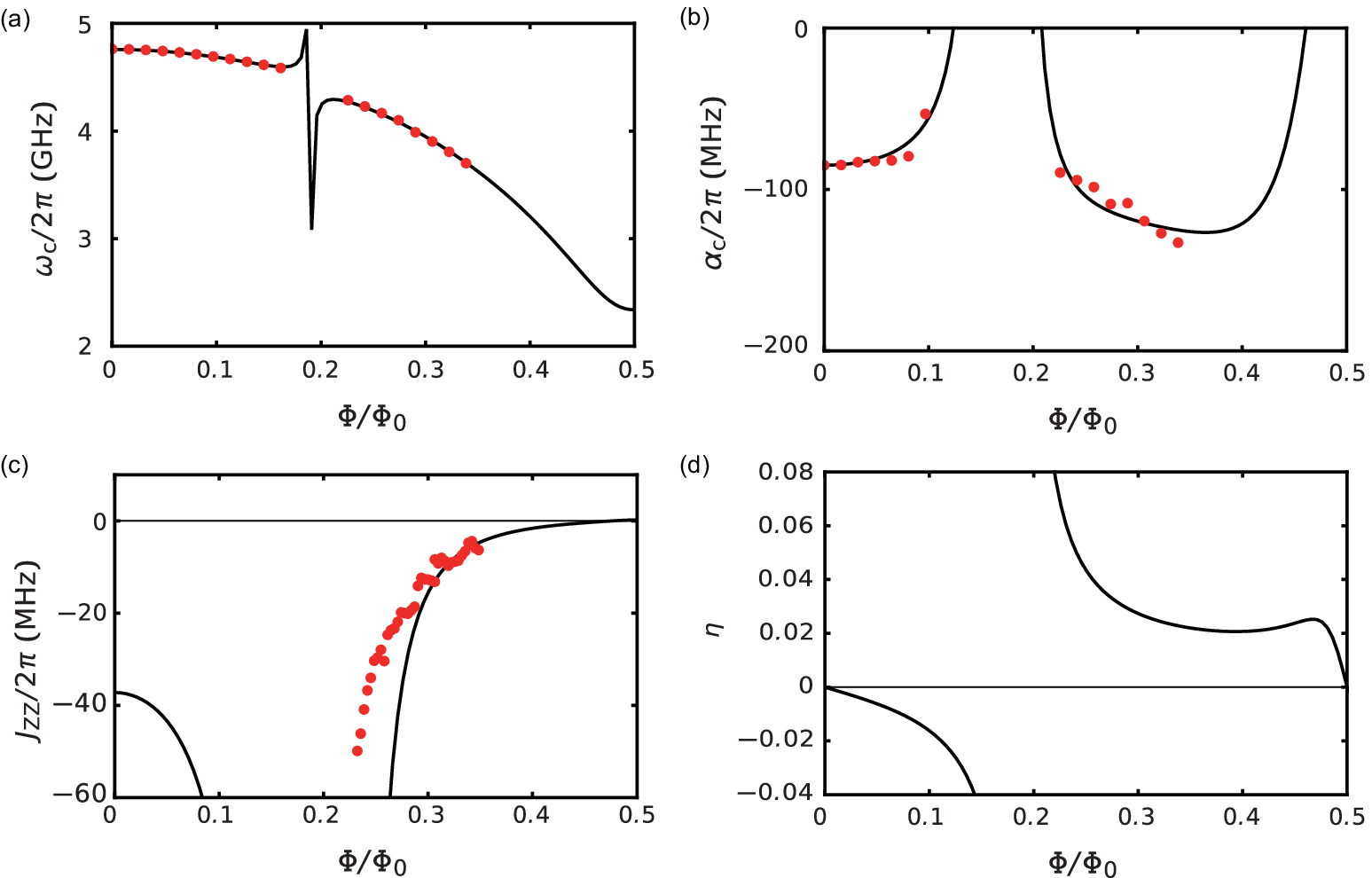}
\caption{
Calibration experiments of the cubic-transmon--transmon coupled system.
(a) Eigenfrequency of the first excited state, $\omega_\mathrm{c}$,  (b) third-order nonlinearity $\alpha _\mathrm{c}$, and (c)~strength of the residual ZZ interaction~$J_\mathrm{ZZ}$ as a function of the flux bias $\Phi$.
Red dots are the experimental data.
(d)~Coupling coefficient of the parametric drive,~$\eta$,  calculated with Eq.~(\ref{eqeta}).
Black curves in (a) and (b) are the fitting results using Eqs.~(\ref{eqomega}) and~(\ref{eqalpha}).
Black curves in (c) and~(d) are calculated based on Eqs.~(\ref{eqzz}) and~(\ref{eqeta}), respectively, with the parameters determined by the fittings.
}
\label{figSfit}
\end{center}
\end{figure}
As described in the main text, the total Hamiltonian $\hat{H}$ of the cubit-transmon--transmon coupled system is given as
\begin{eqnarray}
\hat{H}&=&\hat{H}_0+\hat{V},\\
\hat{H}_0/\hbar &=&\omega _\mathrm{c0}\hat{a}^\dagger\hat{a}+\frac{\alpha_\mathrm{c0}}{2}\hat{a}^\dagger\hat{a}^\dagger\hat{a}\hat{a}
+\omega _\mathrm{t0}\hat{b}^\dagger\hat{b}+\frac{\alpha_\mathrm{t0}}{2}\hat{b}^\dagger\hat{b}^\dagger\hat{b}\hat{b},\\
\hat{V}/\hbar &=&\beta _\mathrm{c0} (\hat{a}^\dagger\hat{a}^\dagger\hat{a}+\hat{a}^\dagger\hat{a}\hat{a})
+g_0(\hat{a}^\dagger\hat{b}+\hat{a}\hat{b}^\dagger ).
\end{eqnarray}
The parameters are defined in the main text.
We treat the off-diagonal part $\hat{V}$ as a perturbative term and obtain the effective Hamiltonian via Schrieffer-Wolff transformation.
\begin{equation}
\hat{H}^\prime \equiv e^{\hat{S}}\hat{H}e^{-\hat{S}}\sim \hat{H}+[\hat{S}, \hat{H}] + \frac{1}{2} [\hat{S}, [\hat{S},\hat{H}]].
\end{equation}
We introduce $\hat{S}_1$ which fulfills
\begin{equation}
\hat{V}=-[\hat{S}_1,\hat{H}_0].
\end{equation}
Then, the effective Hamiltonian in the second order reads
\begin{equation}
\hat{H}_\mathrm{eff}^{(2)}=\hat{H}_0+[\hat{S}_1,\hat{V}]+\frac{1}{2}[\hat{S}_1,[\hat{S}_1,\hat{H}_0]].
\end{equation}

We calculate the effective Hamiltonian by ignoring states with more than four excitation quanta in each qubit and truncating it into a matrix with $16 \times 16$ elements for the two-qubit system.
The calculation is valid when $g_0\ll |\omega _\mathrm{c0}-\omega _\mathrm{t0}|$ and $|\beta _\mathrm{c0}|\ll \omega _\mathrm{c0}$ are satisfied.
The effective Hamiltonian reads~[Eq.~(2) in the main text]
\begin{eqnarray}
\hat{H}_\mathrm{eff}^{(2)}/\hbar =\left[ \omega _\mathrm{c}+g(\hat{b}^\dagger +\hat{b})\right] \hat{a}^\dagger\hat{a}+\frac{\alpha_\mathrm{c}}{2}\hat{a}^\dagger\hat{a}^\dagger\hat{a}\hat{a}
+\omega _\mathrm{t}\hat{b}^\dagger\hat{b}+\frac{\alpha_\mathrm{t}}{2}\hat{b}^\dagger\hat{b}^\dagger\hat{b}\hat{b},\label{eqH}
\end{eqnarray}
where $\omega_\mathrm{c}$, $\omega_\mathrm{t}$, $\alpha_\mathrm{c}$, and $\alpha_\mathrm{t}$ are the eigenmode frequencies and self-Kerr nonlinearities of the qubits after the perturbative treatment of the coupling term. 
They are expressed as follows:
\begin{eqnarray}
\omega _\mathrm{c}\!\!\!&=&\!\!\!\omega _\mathrm{c0}-\frac{2\beta_\mathrm{c0}^2 }{\omega _\mathrm{c0}+\alpha_\mathrm{c0}}+\frac{g_0^2}{\Delta_0},\label{eqomega}\\
\omega _\mathrm{t}\!\!\!&=&\!\!\!\omega _\mathrm{t0}-\frac{g_0^2}{\Delta_0},\\
\alpha _\mathrm{c}\!\!\!&=&\!\!\!\alpha _\mathrm{c0}-\frac{6\beta_\mathrm{c0}^2\omega_\mathrm{c0}}{(\omega _\mathrm{c0}+\alpha_\mathrm{c0})(2\alpha_\mathrm{c0}+\omega _\mathrm{c0})}-\frac{2g_0^2\alpha _\mathrm{c0}}{\Delta_0 (\alpha _\mathrm{c0}+\Delta_0)},\label{eqalpha}\\
\alpha _\mathrm{t}\!\!\!&=&\!\!\!\alpha _\mathrm{t0}+\frac{2g_0^2\alpha _\mathrm{t0}}{\Delta _0(\alpha _\mathrm{t0}-\Delta_0)},
\end{eqnarray}
where $\Delta_0 \equiv \omega_\mathrm{t0} - \omega_\mathrm{c0}$ is the detuning between the qubit bare frequencies.
We use these expressions to fit the experimental data, as shown in Figs.~\ref{figSfit}(a) and~\ref{figSfit}(b).
The fitting parameters are listed in Table~S1. 

The term $g\hat{a}^\dagger\hat{a}(\hat{b}^\dagger +\hat{b}) $ in the effective Hamiltonian~[Eq.~(\ref{eqH})] is the parity-violating term, and the effective couping strength $g$ is expressed as
\begin{equation}
g= -\frac{g_0 \beta_\mathrm{c0}(\Delta _0+\omega _\mathrm{c0} +2\alpha _\mathrm{c0})}{(\alpha_\mathrm{c0}+\Delta _0)(\alpha _\mathrm{c0}+\omega _\mathrm{c0})}.
\end{equation}
The amplitude of the residual ZZ interaction, $J_\mathrm{ZZ}$, between the qubits is derived through Schrieffer-Wolff transformation up to the fourth order of $g_0$,
\begin{eqnarray}
\hat{H}_\mathrm{ZZ}&=&J_\mathrm{ZZ}\hat{a}^\dagger\hat{a}\hat{b}^\dagger\hat{b},\\
J_\mathrm{ZZ}&=&\frac{2 g_0^2 (\alpha _\mathrm{c0}+\alpha _\mathrm{t0})}{(\alpha _\mathrm{c0}+\Delta _0)(-\alpha _\mathrm{t0}+\Delta _0)}.\label{eqzz}
\end{eqnarray}
In Fig.~\ref{figSfit}(c), we plot $J_\mathrm{ZZ}$ obtained with the parameters that are determined from the fittings in Figs.~\ref{figSfit}(a) and~\ref{figSfit}(b).
In the dispersive regime of the two qubits, i.e., for $\Delta _0\gg g_0$, the experimental data in Fig.~\ref{figSfit}(c) has a good agreement with the theoretically expected values. 

For the calculation of the parametric coupling, we continue this procedure one more step.
We set $\hat{S}_2$ such that
\begin{equation}
\hat{V}_2=-[\hat{S}_2,\hat{H}_0],
\end{equation}
where $\hat{V}_2$ is the off-diagonal part of the effective Hamiltonian $\hat{H}_\mathrm{eff}^{(2)}$.
We drive this system at the frequency $\omega_\mathrm{d}$ with a phase $\theta$, such that
\begin{equation}
\hat{H}_\mathrm{d}=\Omega (e^{-i\omega _\mathrm{d} t-i\theta}\hat{a}+e^{i\omega _\mathrm{d} t+i\theta}\hat{a}^\dagger ),
\end{equation} 
and transform the drive Hamiltonian as
\begin{equation}
\hat{H}_\mathrm{dp}=\hat{H}_\mathrm{d}+[\hat{S}_1,\hat{H}_\mathrm{d}]+\frac{1}{2}[\hat{S}_1,[\hat{S}_1,\hat{H}_\mathrm{d}]]+[\hat{S}_2,\hat{H}_\mathrm{d}],
\end{equation} 
to obtain the parametric coupling of the SWAP interaction for $\omega_\mathrm{d} = \Delta \equiv \omega_\mathrm{t} - \omega_\mathrm{c}$~[Eqs.~(3) and~(4) in the main text],
\begin{eqnarray}
\hat{H}_\mathrm{p}/\hbar &=&
\eta\Omega (e^{i\omega _\mathrm{d}t+i\theta}\hat{a}^\dagger \hat{b}+e^{-i\omega _\mathrm{d}t-i\theta}\hat{a} \hat{b}^\dagger ),\\
\eta&\equiv &\frac{-2g_0\beta_\mathrm{c0}(2\omega _\mathrm{c0}^2-\alpha_\mathrm{c0}\Delta _0+ 2\alpha_\mathrm{c0}\omega_\mathrm{c0})}{\Delta _0(\Delta _0-\omega _\mathrm{c0})(\alpha_\mathrm{c0}+\omega _\mathrm{c0})(\alpha_\mathrm{c0}+\omega _\mathrm{c0}+\Delta _0)}.\label{eqeta}
\end{eqnarray}
For the CZ gate, we use the transition at $\omega_\mathrm{d} = \Delta + \alpha_\mathrm{t}$ involving the second-excited state of the transmon, whose amplitude is similarly obtained as
\begin{eqnarray}
\eta _\mathrm{CZ}&\equiv &\frac{2\sqrt{2}g_0\beta_\mathrm{c0}(2\omega _\mathrm{c0}^2-\alpha_\mathrm{c0}\Delta _0+ 2\alpha_\mathrm{c0}\omega_\mathrm{c0}+\alpha_\mathrm{c0}\alpha_\mathrm{t0})}{(\Delta _0-\alpha _\mathrm{t0})(\alpha _\mathrm{t0}-\Delta_0 +\omega _\mathrm{c0})(\alpha_\mathrm{c0}+\omega _\mathrm{c0})(\alpha_\mathrm{c0}-\alpha_\mathrm{t0}+\omega _\mathrm{c0}+\Delta_0 )}.
\end{eqnarray}

\section{PARAMETRICALLY-INDUCED transition}

\begin{figure}[t]
\begin{center}
   \includegraphics[width=8.0cm,angle=0]{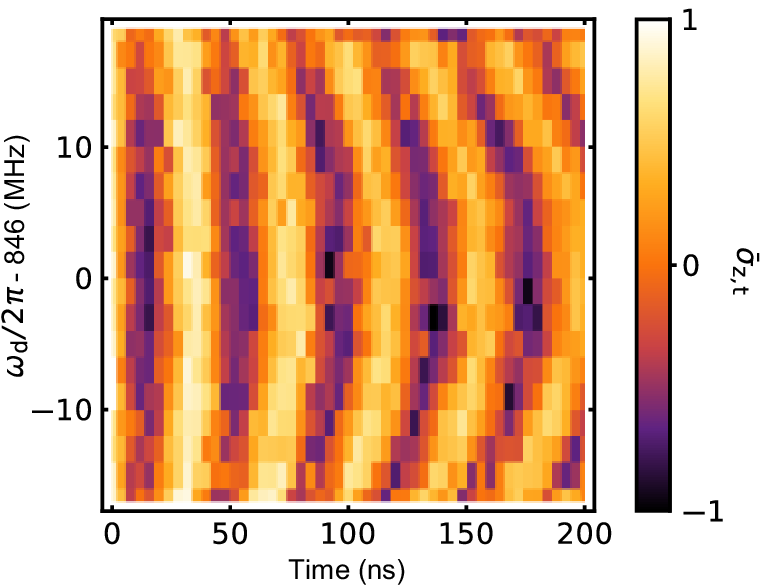}
\caption{
Rabi oscillation of the $\widetilde{|ge\rangle} \leftrightarrow \widetilde{|eg\rangle}$ transition.
The vertical axis is the frequency of the parametric drive, $\omega_\mathrm{d}$, and the horizontal axis is the interaction time.
The color shows the normalized average quadrature amplitude $\bar{\sigma}_{z,\mathrm{t}}$ of the transmon readout signal.
}
\label{figSshev}
\end{center}
\end{figure}

Figure \ref{figSshev} shows the experimental data of the parametrically-induced $\widetilde{|ge\rangle} \leftrightarrow \widetilde{|eg\rangle}$ transition.
We prepare the $\widetilde{|eg\rangle}$ state with a $\pi$-pulse to the cubic transmon and apply the parametric drive to the cubic transmon. 
The excitation is swapped between the two states by the parametric transition.
The resonance frequency, 846 MHz, is the frequency difference between eigenfrequencies of the cubic transmon and transmon.
The Rabi frequency is proportional to the amplitude of the drive, and the maximum Rabi frequency of 30~MHz is obtained.

\section{Randomized Benchmarking}

\begin{figure}[t]
\begin{center}
   \includegraphics[width=12.0cm,angle=0]{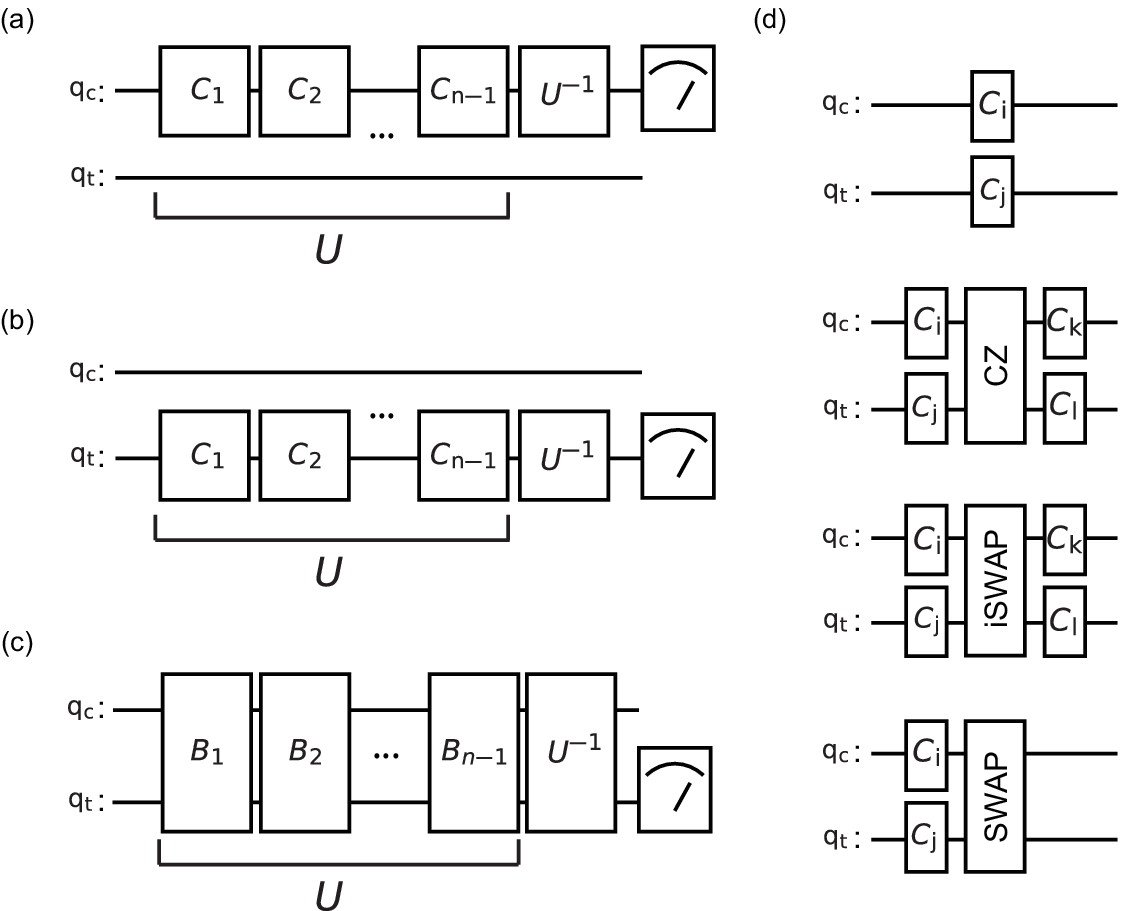}
\caption{
Pulse sequences for RB.
(a) and (b)~Pulse sequences for single-qubit RB with the cubic transmon and transmon, respectively.
An array of single-qubit random Clifford gates, $C_1, C_2, \ldots,  C_{n-1}$, is applied, and $U^{-1}$ is the inverse of the preceding sequence.
(c)~Pulse sequence for two-qubit RB.
An array of two-qubit random Clifford gates, $B_1, B_2, \ldots,  B_{n-1}$, is applied, followed by the inverse $U^{-1}$.
(d) Decompositions of two-qubit Clifford gates.
}
\label{figSrbpulse}
\end{center}
\end{figure}

\begin{figure}[t]
\begin{center}
   \includegraphics[width=15.0cm,angle=0]{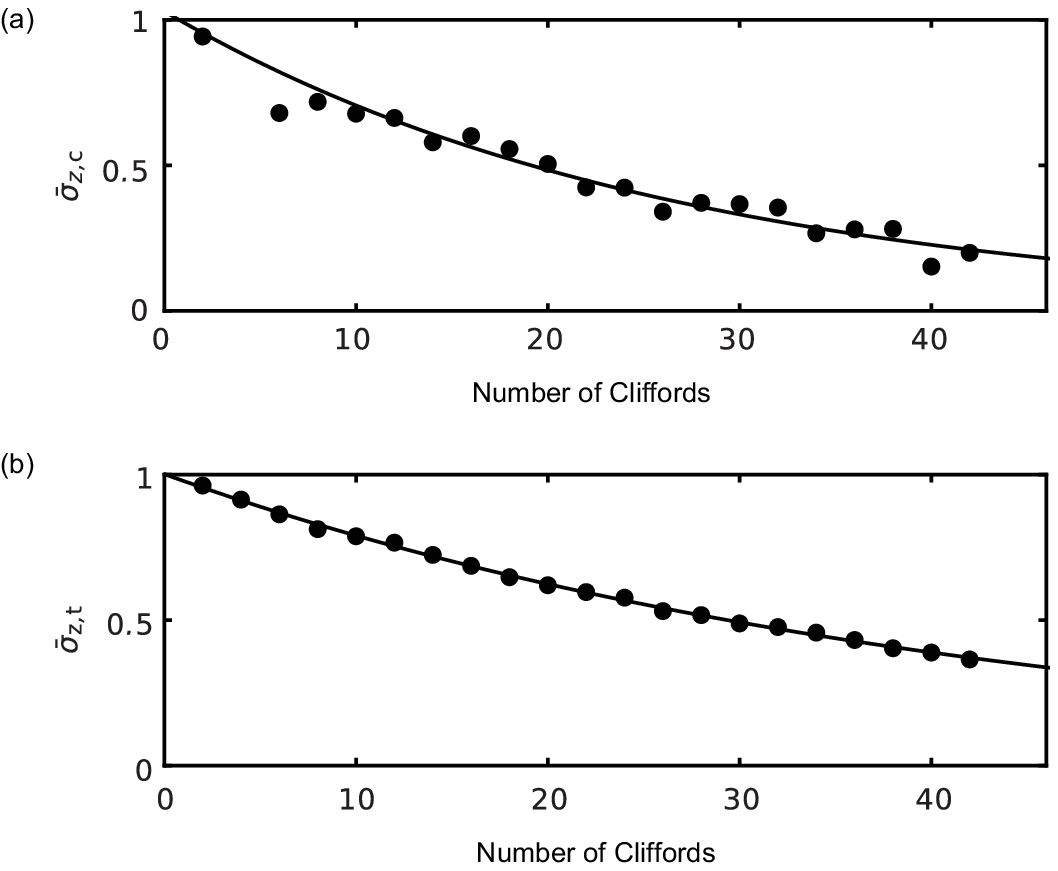}
\caption{
Randomized benchmarking of the single-qubit gates.
The vertical axes show the normalized average quadrature amplitudes of the readout signals for (a)~the cubic transmon and (b)~the transmon.
The horizontal axes show the number of Clifford gates applied in the randomized sequence. Curves are the fittings to the depolarization model.
}
\label{figSrb1}
\end{center}
\end{figure}

Figure \ref{figSrbpulse} shows the gate sequences for the single-qubit and two-qubit randomized benchmarking~(RB). 
We drive the cubic transmon with a CW field to eliminate the static ZZ interaction~(not shown).
The pulse shapes for the single-qubit gates are Gaussian with a full width at half maximum of 18.6~ns.
The swap pulse and each segment of the control-phase pulse [Fig.~3(a) in the main text] have rising and falling edges of a Gaussian shape with the half width at half maximum of 3.0~ns and 1.5~ns, respectively.
The length of the flat-top region is 32~ns for the swap pulse and 16~ns for each segment of the control-phase pulse.

The tails for all pulse are truncated when the amplitudes become $10^{-3}$ times smaller than the maximum. 
For the interleaved RB, we add a target gate (CZ, iSWAP and SWAP) between each Clifford gates. We repeat the sequences 5000  times with 100 (50)
different random patterns for the protocol in Figs.~\ref{figSrbpulse}(a) and~\ref{figSrbpulse}(b) [Fig.~\ref{figSrbpulse}(c)]. 
We measure the average value of the $\sigma _z$-component of the cubic transmon in the protocol in Fig.~\ref{figSrbpulse}(a) and that of the transmon in Figs.~\ref{figSrbpulse}(b) and~\ref{figSrbpulse}(c).

Figure~\ref{figSrb1} shows the results of standard RB for the single-qubit gates. 
The average gate fidelities of the single-qubit gates are evaluated to be $0.963\pm 0.001$ for the cubic transmon and $0.977\pm 0.001$ for the transmon.

\section{Raman transition through a CONTINUOUS MICROWAVE field}

\begin{figure}[h]
\begin{center}
   \includegraphics[width=8.0cm,angle=0]{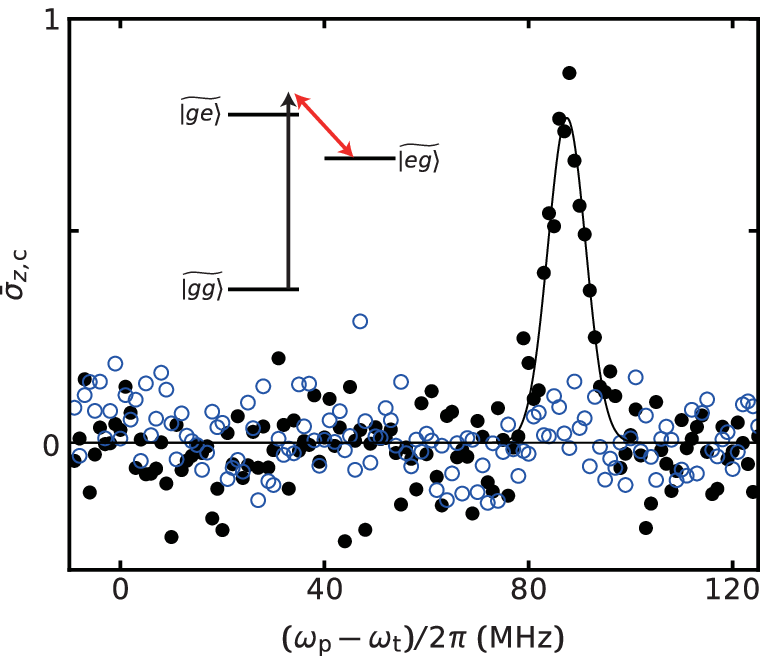}
\caption{
Raman transition assisted by the CW drive. 
The vertical axis shows the normalized average quadrature amplitude of the readout signal for the cubic transmon.
Horizontal axis is the frequency of the probe pulse, $\omega_\mathrm{p}$, applied to the transmon, subtracted by the eigenfrequency of the transmon, $\omega _\mathrm{t}$.
Black dots and blue circles are the experimental data with and without the CW drive. 
The black curve fits the data with a Gaussian function, whose spectral width is one of the fitting parameters and consistent with the temporal shape of the probe pulse.
Inset shows the energy-level diagram. Red and black arrows represent the CW-drive and probe-microwave frequencies, respectively.
}
\label{figSraman}
\end{center}
\end{figure}

In the main text, we irradiate the cubic transmon with a continuous microwave~(CW) drive to eliminate the unwanted static ZZ interaction between the two qubits.
However, this CW drive also induces an unwanted Raman transition which is mediated by the transmon excitation.
Figure~\ref{figSraman} shows the experimental data regarding the transition with pulsed spectroscopy.
We sweep the frequency of the probe microwave pulse, $\omega_\mathrm{p}$, around the transmon excitation frequency and measure the state of the cubic transmon.
The pulse has a Gaussian shape with the full width at half maximum of 60~ns.
The peak observed in Fig.~\ref{figSraman} corresponds to the Raman transition process depicted in the inset.
In accordance with the CW-drive detuning of 84~MHz from the $\widetilde{|eg\rangle}\leftrightarrow\widetilde{|ge\rangle}$ transition, the Raman transition appears at $(\omega_\mathrm{t}/2\pi + 84)$~MHz.
This transition is close to the transmon resonance and can be an error source for single-qubit gates with a short pulse.
This error can be suppressed by the use of DRAG pulses~\cite{DRAG}.


\begin{thebibliography}{99} 

\bibitem{martinis2014}R. Barends \textit{et al.,} Nature \textbf{508}, 500 (2014).
\bibitem{martinis2019}F. Arute \textit{et al.,} Nature \textbf{574}, 505 (2019).
\bibitem{johnson2020}S. S. Hong \textit{et al.,} Phys. Rev. A \textbf{101}, 012302 (2020).
\bibitem{rigetti}S. A. Caldwell \textit{et al.,} Phys. Rev. Appl. \textbf{10}, 034050 (2018).
\bibitem{ibm2016}S. Sheldon, E. Magesan, J. M. Chow, and J. M. Gambetta, Phys. Rev. A \textbf{93}, 060302(R) (2016).

\bibitem{schoelkopf2012}M. D. Reed \textit{et al.,} Nature \textbf{482}, 382 (2012). 
\bibitem{martinis2015}J. Kelly \textit{et al.,} Nature \textbf{519}, 66 (2015).
\bibitem{schoelkopf2016}N. Ofek \textit{et al.,} Nature \textbf{536}, 441 (2016).
\bibitem{correct2019}L. Hu \textit{et al.,} Nat. Phys. \textbf{15}, 503 (2019).
\bibitem{wallraf2019}C. K. Andersen \textit{et al.,} npj Quantum Inf. \textbf{5}, 69 (2019).

\bibitem{fowler}A. G. Fowler, M. Mariantoni, J. M. Martinis, and A. N. Cleland, Phys. Rev. A \textbf{86}, 032324 (2012). 
\bibitem{terhal}B. M. Terhal \textit{et al}., arXiv:2002.11008 (2002).

\bibitem{nakamura2007}A. O. Niskanen \textit{et al.,} Science \textbf{316}, 723 (2007).
\bibitem{ibm2011}J. M. Chow \textit{et al.,} Phys. Rev. Lett. \textbf{107}, 080502 (2011).
\bibitem{screview}P. Krantz \textit{et al.,} Appl. Phys. Rev. \textbf{6}, 021318 (2019).

\bibitem{DiVincenzo2017}S. Richer, N. Maleeva, S. T. Skacel, I. M. Pop, D. DiVincenzo, Phys. Rev. B \textbf{96}, 174520 (2017).
\bibitem{steele2018}M. Kounalakis, C. Dickel, A. Bruno, N. K. Langford, and G. A. Steele, npj Quantum Information \textbf{4}, 38 (2018).
\bibitem{houck2019}P. Mundada, G. Zhang, T. Hazard, and A. Houck, Phys. Rev. Appl. \textbf{12}, 054023 (2019).
\bibitem{duan2020}X. Han, T. Cai, X. Li, Y. Wu, Y. Ma, J. Wang, H. Zhang, Y. Song, and L. Duan, arXiv:2003.08542 (2020).

\bibitem{plourde2020}J. Ku \textit{et al.,} arXiv:2003.02775 (2020). 

\bibitem{SNAIL} N. E. Frattini, U. Vool, S. Shankar, A. Narla, K. M. Sliwa and M. H. Devoret, Appl. Phys. Lett. \textbf{110}, 222603 (2017).

\bibitem{SNAIL2} N. E. Frattini, V. V. Sivak, A. Lingenfelter, S. Shankar, and M. H. Devoret, Phys. Rev. Applied \textbf{10}, 054020 (2018).
\bibitem{SNAIL3} V. V. Sivak, N. E. Frattini, V. R. Joshi, A. Lingenfelter, S. Shankar, and M. H. Devoret,
Phys. Rev. Applied \textbf{11}, 054060 (2019).
\bibitem{noguchi2018}A. Noguchi, R. Yamazaki, Y. Tabuchi, and Y. Nakamura, Nat. Commun. \textbf{11}, 1183 (2020).

\bibitem{nakamura2015}P.-M. Billangeon, J. S. Tsai, and Y. Nakamura, Phys. Rev. B \textbf{91}, 094517 (2015).
\bibitem{DiVincenzo2016}S. Richer and D. DiVincenzo, Phys. Rev. B \textbf{93}, 134501 (2016).
\bibitem{sideband}F. Beaudoin, M. P. daSilva, Z. Dutton, A. Blais, \textit{et al.,} Phys. Rev. A \textbf{86}, 022305 (2012).

\bibitem{koch2007}J. Koch, T. M. Yu, J. Gambetta, A. A. Houck, D. I. Schuster, J. Majer, A. Blais, M. H. Devoret, S. M. Girvin, and R. J. Schoelkopf, Phys. Rev A, \textbf{76}, 042319 (2007).

\bibitem{optomechanics}M. Aspelmeyer, T. J. Kippenberg and F. Marquardt, Rev. Mod. Phys. \textbf{86}, 1391 (2014).
\bibitem{blatt2008} H. H\"{a}ffner, C. F. Roos, and R. Blatt, Physics Reports \textbf{469}, 155 (2008).
\bibitem{ion}F. Mintert and C. Wunderlich, Phys. Rev. Lett. \textbf{87}, 257904 (2001).

\bibitem{supple}\textit{Supplementary Material}

\bibitem{ibm2019}M. Ware \textit{et al.,} arXiv:1905.11480 (2019).
\bibitem{fermi2018}I. D. Kivlichan, J. McClean, N. Wiebe, C. Gidney, A. Aspuru-Guzik, Garnet Kin-Lic Chan, and R. Babbush, Phys. Rev. Lett. \textbf{120}, 110501 (2018).
\bibitem{martinis2020} B. Foxen \textit{et al.,} arXiv:2001.08343 (2020).

\bibitem{RBth1}E. Magesan, J. M. Gambetta, J. Emerson, Phys. Rev. Lett. \textbf{106}, 180504 (2011).
\bibitem{RBth2}E. Magesan \textit{et al.}, Phys. Rev. Lett. \textbf{109}, 080505 (2012).
\bibitem{RB1}A. D. C\'{o}rcoles, J. M. Gambetta, J. M. Chow, J. A. Smolin, M. Ware, J. Strand, B. L. T. Plourde, M. Steffen, Phys. Rev. A \textbf{87}, 030301(R) (2013).
\bibitem{RB2}A. D. C\'{o}rcoles \textit{et al.,} Nat. Commun. \textbf{6}, 6979 (2015).

\end{thebibliography}

\begin{thebibliography}{99} 

\bibitem{DRAG}F. Motzoi, J. M. Gambetta, P. Rebentrost, and F. K. Wilhelm, Phys. Rev. Lett. \textbf{103}, 110501 (2009).

\end{thebibliography}
\end{document}